\documentstyle[epic,eepic]{EuroPhys}

\def\And{{\rm and\ }}

\def\stars{\bigskip\centerline{***}\medskip}

\newif\ifboo \boofalse

\def\Review#1{\boofalse{\it #1},}
\def\Name#1{{\sc #1},}
\def\Vol#1{\ifboo Vol. {\bf #1}\else{\bf #1}\fi}
\def\Year#1{\ifboo #1\else(#1)\fi}
\def\Book#1{\bootrue{\it #1},}
\def\Page#1{\ifboo {\rm p. #1}\else{\rm #1}\fi}

  \newcommand{\epl}{true}
\begin{document}
\euro{40}{6}{625-630}{1997}
\Date{15 December 1997}
\shorttitle{G.H. RISTOW: CRITICAL EXPONENTS FOR GRANULAR PHASE TRANSITIONS}
\title{Critical exponents for granular phase transitions}
\author{Gerald H. Ristow}
\institute{Fachbereich Physik, Philipps--Universit\"at\\
           Renthof 6, D--35032 Marburg, Germany}
\rec{28 February 1997}{in final form 31 October 1997}
\pacs{
\Pacs{46}{10$+$z}{Mechanics of discrete systems}
\Pacs{81}{05Rm}{Porous materials; granular materials}
\Pacs{02}{70Ns}{Molecular dynamics and particle methods}
     }

\maketitle

\begin{abstract}
The solid--fluid phase transition of a granular material shaken horizontally is
investigated numerically. We find that it is a second-order phase transition
and propose two order parameters, namely the averaged kinetic energy and the
averaged granular temperature, to determine the fluidization point precisely.
It scales with the acceleration of the external vibration. Using this
fluidization point as critical point, we discuss the scaling of the kinetic
energy and show that the kinetic energy and the granular temperature show two
different universal critical point exponents for a wide range of excitation
amplitudes.
\end{abstract}
\vspace{-0.75cm}
Granular materials show a fascinating behaviour under a variety of experimental
conditions~\cite{jaeger}. Most striking are the different regimes found under
vertical shaking: for low excitations sound waves are found, for medium
excitations heap formations and convection rolls are visible and for high
excitations period doubling can occur which can lead to standing and travelling
waves~\cite{warr}. This behaviour is connected to two visible phase transitions
where the strength of the excitation is measured in the dimensionless variable
$\Gamma := A_0\, \omega/g$, with $A_0$ and $\omega$ standing for the amplitude
and the angular frequency of the external shaking and $g$ denoting the
gravitational constant. The first transition occurs around
$\Gamma=1$~\cite{evesque,brennen} and the second was found to occur around
$\Gamma=1.8$ in a one-layer system~\cite{brennen}. In a quasi two-dimensional
box, the transition points were significantly higher~\cite{goldshtein}. In the
intermediate regime, the bed expansion scales as $\Gamma$~\cite{brennen},
whereas the supercritical scaling is correlated with $A_0\,
\omega$~\cite{warr,brennen,luding,lee}. In this letter, we will focus on a
slightly different setup and study the first phase transition from a solid-like
to a fluid-like regime of a granular material undergoing {\em horizontal}
vibrations. Our aim will be to quantify this transition point and to show the
correct scaling with $A_0$ and $\omega$. Furthermore, we demonstrate that the
averaged kinetic energy and the granular temperature show beautiful scaling
over more than one decade with different critical exponents.

The system in mind was first described by Stra\ss{}burger et
al.~\cite{strassburger} and an enhanced setup was used in~\cite{ristow} for
quantitative measurements which we will sketch briefly. A few layers of
commercially available glass ballotonies were filled into a 100 mm long and 0.6
mm wide channel and exposed to horizontal vibrations of the form $A(t) = A_0
\sin(\omega t)$ where $\omega=2\pi f$. In order to get better defined surface
properties for particle-wall contacts a monolayer of smaller spheres was glued
onto the channel bottom. Since the channel width is equal to the maximum
particle diameter, one gets a quasi two-dimensional system which can easily be
observed from the side. For low excitations, either low frequencies $f$ or
small amplitudes $A_0$, no relative motion of particles with respect to the
container is visible and the spheres form a nearly perfect triangular lattice,
sketched in fig.~\ref{fig: 1}a. For higher excitations, relative motion sets in
and the lattice breaks up, sketched in  fig.~\ref{fig: 1}b. Using this system,
the phase diagram for the solid-fluid transition was investigated and it was
found that the transition point scales with $\Gamma$ using experiments and
numerical simulations~\cite{ristow}. A summary of the phase diagram is shown in
fig.~\ref{fig: 2}, demonstrating that the numerical model accurately describes
the experiment and the slight shift could be attributed to the higher wall
friction present in the experiment. For the experimental values, the onset of
grain motion is denoted by $f_0$ and the extrapolated critical frequency by
$f_c$.
\ifx\epl\undefined
\begin{figure}[ht]
  \centerline{ 
\ifx\epl\undefined
  \setlength{\unitlength}{0.221pt}
  \begin{picture}(1049,900)(0,0)
\else
  \setlength{\unitlength}{0.2pt}
  \begin{picture}(963,825)(-60,-80)
\fi

\put(200,625){\circle{100}}
\put(300,625){\circle{100}}
\put(400,625){\circle{100}}
\put(500,625){\circle{100}}
\put(600,625){\circle{100}}
\put(700,625){\circle{100}}
\put(800,625){\circle{100}}
\put(900,625){\circle{100}}
\put(250,710){\circle{100}}
\put(350,710){\circle{100}}
\put(450,710){\circle{100}}
\put(550,710){\circle{100}}
\put(650,710){\circle{100}}
\put(750,710){\circle{100}}
\put(850,710){\circle{100}}
\put(950,710){\circle{100}}
\put(200,550){\circle*{50}}
\put(250,550){\circle*{50}}
\put(300,550){\circle*{50}}
\put(350,550){\circle*{50}}
\put(400,550){\circle*{50}}
\put(450,550){\circle*{50}}
\put(500,550){\circle*{50}}
\put(550,550){\circle*{50}}
\put(600,550){\circle*{50}}
\put(650,550){\circle*{50}}
\put(700,550){\circle*{50}}
\put(750,550){\circle*{50}}
\put(800,550){\circle*{50}}
\put(850,550){\circle*{50}}
\put(900,550){\circle*{50}}
\put(950,550){\circle*{50}}

\put(200,140){\circle{100}}
\put(310,145){\circle{100}}
\put(415,135){\circle{100}}
\put(515,140){\circle{100}}
\put(625,145){\circle{100}}
\put(740,145){\circle{100}}
\put(845,135){\circle{100}}
\put(960,140){\circle{100}}
\put(230,250){\circle{100}}
\put(350,255){\circle{100}}
\put(465,260){\circle{100}}
\put(590,255){\circle{100}}
\put(710,260){\circle{100}}
\put(820,245){\circle{100}}
\put(935,250){\circle{100}}
\put(200,50){\circle*{50}}
\put(250,50){\circle*{50}}
\put(300,50){\circle*{50}}
\put(350,50){\circle*{50}}
\put(400,50){\circle*{50}}
\put(450,50){\circle*{50}}
\put(500,50){\circle*{50}}
\put(550,50){\circle*{50}}
\put(600,50){\circle*{50}}
\put(650,50){\circle*{50}}
\put(700,50){\circle*{50}}
\put(750,50){\circle*{50}}
\put(800,50){\circle*{50}}
\put(850,50){\circle*{50}}
\put(900,50){\circle*{50}}
\put(950,50){\circle*{50}}

\put(600,840){\makebox(0,0){\bf (a)}}
\put(600,380){\makebox(0,0){\bf (b)}}

\end{picture} }
  \caption{Sketch of the physical system undergoing horizontal vibrations for 
           (a) low and (b) higher excitations.}
  \label{fig: 1}
\end{figure}
\begin{figure}[ht]
  \centerline{ 
\ifx\epl\undefined
  \setlength{\unitlength}{0.240900pt}
  \begin{picture}(963,825)(0,0)
\else
  \setlength{\unitlength}{0.22pt}
  \begin{picture}(963,825)(30,0)
\fi
\thicklines \path(220,113)(240,113)
\thicklines \path(985,113)(965,113)
\put(198,113){\makebox(0,0)[r]{0}}
\thicklines \path(220,266)(240,266)
\thicklines \path(985,266)(965,266)
\put(198,266){\makebox(0,0)[r]{1}}
\thicklines \path(220,419)(240,419)
\thicklines \path(985,419)(965,419)
\put(198,419){\makebox(0,0)[r]{2}}
\thicklines \path(220,571)(240,571)
\thicklines \path(985,571)(965,571)
\put(198,571){\makebox(0,0)[r]{3}}
\thicklines \path(220,724)(240,724)
\thicklines \path(985,724)(965,724)
\put(198,724){\makebox(0,0)[r]{4}}
\thicklines \path(220,877)(240,877)
\thicklines \path(985,877)(965,877)
\put(198,877){\makebox(0,0)[r]{5}}
\thicklines \path(220,113)(220,133)
\thicklines \path(220,877)(220,857)
\put(220,68){\makebox(0,0){0}}
\thicklines \path(348,113)(348,133)
\thicklines \path(348,877)(348,857)
\put(348,68){\makebox(0,0){1}}
\thicklines \path(475,113)(475,133)
\thicklines \path(475,877)(475,857)
\put(475,68){\makebox(0,0){2}}
\thicklines \path(603,113)(603,133)
\thicklines \path(603,877)(603,857)
\put(603,68){\makebox(0,0){3}}
\thicklines \path(730,113)(730,133)
\thicklines \path(730,877)(730,857)
\put(730,68){\makebox(0,0){4}}
\thicklines \path(858,113)(858,133)
\thicklines \path(858,877)(858,857)
\put(858,68){\makebox(0,0){5}}
\thicklines \path(985,113)(985,133)
\thicklines \path(985,877)(985,857)
\put(985,68){\makebox(0,0){6}}
\thicklines \path(220,113)(985,113)(985,877)(220,877)(220,113)
\put(45,495){\makebox(0,0)[l]{\shortstack{$f$ \\ \protect{[Hz]}}}}
\put(602,0){\makebox(0,0){$A_0$ [cm]}}
\put(348,266){\makebox(0,0)[l]{SOLID}}
\put(666,495){\makebox(0,0)[l]{FLUID}}
\put(858,801){\makebox(0,0)[r]{$f_0$ (exp.)}}
\put(902,801){\makebox(0,0){$\star$}}
\put(373,576){\makebox(0,0){$\star$}}
\put(437,477){\makebox(0,0){$\star$}}
\put(475,474){\makebox(0,0){$\star$}}
\put(297,648){\makebox(0,0){$\star$}}
\put(373,480){\makebox(0,0){$\star$}}
\put(475,419){\makebox(0,0){$\star$}}
\put(526,357){\makebox(0,0){$\star$}}
\put(603,342){\makebox(0,0){$\star$}}
\put(794,296){\makebox(0,0){$\star$}}
\put(858,756){\makebox(0,0)[r]{$f_c$ (exp.)}}
\put(902,756){\circle*{12}}
\put(373,637){\circle*{12}}
\put(437,544){\circle*{12}}
\put(475,556){\circle*{12}}
\put(297,739){\circle*{12}}
\put(373,648){\circle*{12}}
\put(475,533){\circle*{12}}
\put(526,442){\circle*{12}}
\put(603,426){\circle*{12}}
\put(794,380){\circle*{12}}
\put(858,711){\makebox(0,0)[r]{simulation}}
\put(902,711){\circle{18}}
\put(284,648){\circle{18}}
\put(316,556){\circle{18}}
\put(348,518){\circle{18}}
\put(379,464){\circle{18}}
\put(411,419){\circle{18}}
\put(475,373){\circle{18}}
\put(603,327){\circle{18}}
\put(730,293){\circle{18}}
\put(794,283){\circle{18}}
\put(858,278){\circle{18}}
\put(985,266){\circle{18}}
\thinlines \path(880,711)(946,711)
\thinlines \path(880,721)(880,701)
\thinlines \path(946,721)(946,701)
\thinlines \path(284,571)(284,724)
\thinlines \path(274,571)(294,571)
\thinlines \path(274,724)(294,724)
\thinlines \path(316,510)(316,602)
\thinlines \path(306,510)(326,510)
\thinlines \path(306,602)(326,602)
\thinlines \path(348,487)(348,548)
\thinlines \path(338,487)(358,487)
\thinlines \path(338,548)(358,548)
\thinlines \path(379,449)(379,480)
\thinlines \path(369,449)(389,449)
\thinlines \path(369,480)(389,480)
\thinlines \path(411,411)(411,426)
\thinlines \path(401,411)(421,411)
\thinlines \path(401,426)(421,426)
\thinlines \path(475,357)(475,388)
\thinlines \path(465,357)(485,357)
\thinlines \path(465,388)(485,388)
\thinlines \path(603,319)(603,335)
\thinlines \path(593,319)(613,319)
\thinlines \path(593,335)(613,335)
\thinlines \path(730,285)(730,300)
\thinlines \path(720,285)(740,285)
\thinlines \path(720,300)(740,300)
\thinlines \path(794,278)(794,287)
\thinlines \path(784,278)(804,278)
\thinlines \path(784,287)(804,287)
\thinlines \path(858,274)(858,282)
\thinlines \path(848,274)(868,274)
\thinlines \path(848,282)(868,282)
\thinlines \path(985,264)(985,267)
\thinlines \path(975,264)(995,264)
\thinlines \path(975,267)(995,267)
\thinlines \path(284,642)(284,642)(291,615)(298,592)(305,571)(312,554)(319,537)(326,523)(333,510)(340,498)(348,487)(355,477)(362,468)(369,460)(376,452)(383,444)(390,437)(397,431)(404,424)(411,419)(418,413)(425,408)(433,403)(440,398)(447,394)(454,389)(461,385)(468,381)(475,378)(482,374)(489,371)(496,367)(503,364)(510,361)(518,358)(525,355)(532,352)(539,350)(546,347)(553,345)(560,342)(567,340)(574,338)(581,335)(588,333)(595,331)(603,329)(610,327)(617,325)(624,323)(631,322)
\thinlines \path(631,322)(638,320)(645,318)(652,316)(659,315)(666,313)(673,312)(680,310)(688,309)(695,307)(702,306)(709,304)(716,303)(723,301)(730,300)(737,299)(744,298)(751,296)(758,295)(765,294)(773,293)(780,292)(787,291)(794,289)(801,288)(808,287)(815,286)(822,285)(829,284)(836,283)(843,282)(850,281)(858,280)(865,279)(872,279)(879,278)(886,277)(893,276)(900,275)(907,274)(914,273)(921,273)(928,272)(935,271)(942,270)(950,269)(957,269)(964,268)(971,267)(978,267)(985,266)
\thinlines \drawline[-95](293,877)(298,853)(305,822)(312,794)(319,769)(326,747)(333,727)(340,709)(348,692)(355,676)(362,662)(369,649)(376,637)(383,625)(390,614)(397,604)(404,595)(411,586)(418,577)(425,569)(433,561)(440,554)(447,547)(454,540)(461,534)(468,528)(475,522)(482,517)(489,511)(496,506)(503,501)(510,496)(518,492)(525,487)(532,483)(539,479)(546,475)(553,471)(560,467)(567,464)(574,460)(581,457)(588,454)(595,450)(603,447)(610,444)(617,441)(624,438)(631,435)(638,433)(645,430)
\thinlines \drawline[-95](645,430)(652,427)(659,425)(666,422)(673,420)(680,418)(688,415)(695,413)(702,411)(709,409)(716,406)(723,404)(730,402)(737,400)(744,398)(751,397)(758,395)(765,393)(773,391)(780,389)(787,388)(794,386)(801,384)(808,383)(815,381)(822,379)(829,378)(836,376)(843,375)(850,373)(858,372)(865,370)(872,369)(879,368)(886,366)(893,365)(900,364)(907,362)(914,361)(921,360)(928,359)(935,357)(942,356)(950,355)(957,354)(964,353)(971,351)(978,350)(985,349)
\end{picture} }
  \caption{Phase diagram for the solid--fluid transition given in 
    ref.~\protect{\cite{ristow}}: experimental points are $\star\ (f_0)$ and 
    $\bullet\ (f_c)$; data from numerical simulations are $\circ$ 
    with solid line as best fit. Dashed line shows theoretical curve for
    equal size particles.}
  \label{fig: 2}
\end{figure}
\else
\begin{figure}[ht]
  { \hspace{-1.5cm}
\ifx\epl\undefined
  \setlength{\unitlength}{0.221pt}
  \begin{picture}(1049,900)(0,0)
\else
  \setlength{\unitlength}{0.2pt}
  \begin{picture}(963,825)(-60,-80)
\fi

\put(200,625){\circle{100}}
\put(300,625){\circle{100}}
\put(400,625){\circle{100}}
\put(500,625){\circle{100}}
\put(600,625){\circle{100}}
\put(700,625){\circle{100}}
\put(800,625){\circle{100}}
\put(900,625){\circle{100}}
\put(250,710){\circle{100}}
\put(350,710){\circle{100}}
\put(450,710){\circle{100}}
\put(550,710){\circle{100}}
\put(650,710){\circle{100}}
\put(750,710){\circle{100}}
\put(850,710){\circle{100}}
\put(950,710){\circle{100}}
\put(200,550){\circle*{50}}
\put(250,550){\circle*{50}}
\put(300,550){\circle*{50}}
\put(350,550){\circle*{50}}
\put(400,550){\circle*{50}}
\put(450,550){\circle*{50}}
\put(500,550){\circle*{50}}
\put(550,550){\circle*{50}}
\put(600,550){\circle*{50}}
\put(650,550){\circle*{50}}
\put(700,550){\circle*{50}}
\put(750,550){\circle*{50}}
\put(800,550){\circle*{50}}
\put(850,550){\circle*{50}}
\put(900,550){\circle*{50}}
\put(950,550){\circle*{50}}

\put(200,140){\circle{100}}
\put(310,145){\circle{100}}
\put(415,135){\circle{100}}
\put(515,140){\circle{100}}
\put(625,145){\circle{100}}
\put(740,145){\circle{100}}
\put(845,135){\circle{100}}
\put(960,140){\circle{100}}
\put(230,250){\circle{100}}
\put(350,255){\circle{100}}
\put(465,260){\circle{100}}
\put(590,255){\circle{100}}
\put(710,260){\circle{100}}
\put(820,245){\circle{100}}
\put(935,250){\circle{100}}
\put(200,50){\circle*{50}}
\put(250,50){\circle*{50}}
\put(300,50){\circle*{50}}
\put(350,50){\circle*{50}}
\put(400,50){\circle*{50}}
\put(450,50){\circle*{50}}
\put(500,50){\circle*{50}}
\put(550,50){\circle*{50}}
\put(600,50){\circle*{50}}
\put(650,50){\circle*{50}}
\put(700,50){\circle*{50}}
\put(750,50){\circle*{50}}
\put(800,50){\circle*{50}}
\put(850,50){\circle*{50}}
\put(900,50){\circle*{50}}
\put(950,50){\circle*{50}}

\put(600,840){\makebox(0,0){\bf (a)}}
\put(600,380){\makebox(0,0){\bf (b)}}

\end{picture} \hfill \hspace{-0.5cm}
\ifx\epl\undefined
  \setlength{\unitlength}{0.240900pt}
  \begin{picture}(963,825)(0,0)
\else
  \setlength{\unitlength}{0.22pt}
  \begin{picture}(963,825)(30,0)
\fi
\thicklines \path(220,113)(240,113)
\thicklines \path(985,113)(965,113)
\put(198,113){\makebox(0,0)[r]{0}}
\thicklines \path(220,266)(240,266)
\thicklines \path(985,266)(965,266)
\put(198,266){\makebox(0,0)[r]{1}}
\thicklines \path(220,419)(240,419)
\thicklines \path(985,419)(965,419)
\put(198,419){\makebox(0,0)[r]{2}}
\thicklines \path(220,571)(240,571)
\thicklines \path(985,571)(965,571)
\put(198,571){\makebox(0,0)[r]{3}}
\thicklines \path(220,724)(240,724)
\thicklines \path(985,724)(965,724)
\put(198,724){\makebox(0,0)[r]{4}}
\thicklines \path(220,877)(240,877)
\thicklines \path(985,877)(965,877)
\put(198,877){\makebox(0,0)[r]{5}}
\thicklines \path(220,113)(220,133)
\thicklines \path(220,877)(220,857)
\put(220,68){\makebox(0,0){0}}
\thicklines \path(348,113)(348,133)
\thicklines \path(348,877)(348,857)
\put(348,68){\makebox(0,0){1}}
\thicklines \path(475,113)(475,133)
\thicklines \path(475,877)(475,857)
\put(475,68){\makebox(0,0){2}}
\thicklines \path(603,113)(603,133)
\thicklines \path(603,877)(603,857)
\put(603,68){\makebox(0,0){3}}
\thicklines \path(730,113)(730,133)
\thicklines \path(730,877)(730,857)
\put(730,68){\makebox(0,0){4}}
\thicklines \path(858,113)(858,133)
\thicklines \path(858,877)(858,857)
\put(858,68){\makebox(0,0){5}}
\thicklines \path(985,113)(985,133)
\thicklines \path(985,877)(985,857)
\put(985,68){\makebox(0,0){6}}
\thicklines \path(220,113)(985,113)(985,877)(220,877)(220,113)
\put(45,495){\makebox(0,0)[l]{\shortstack{$f$ \\ \protect{[Hz]}}}}
\put(602,0){\makebox(0,0){$A_0$ [cm]}}
\put(348,266){\makebox(0,0)[l]{SOLID}}
\put(666,495){\makebox(0,0)[l]{FLUID}}
\put(858,801){\makebox(0,0)[r]{$f_0$ (exp.)}}
\put(902,801){\makebox(0,0){$\star$}}
\put(373,576){\makebox(0,0){$\star$}}
\put(437,477){\makebox(0,0){$\star$}}
\put(475,474){\makebox(0,0){$\star$}}
\put(297,648){\makebox(0,0){$\star$}}
\put(373,480){\makebox(0,0){$\star$}}
\put(475,419){\makebox(0,0){$\star$}}
\put(526,357){\makebox(0,0){$\star$}}
\put(603,342){\makebox(0,0){$\star$}}
\put(794,296){\makebox(0,0){$\star$}}
\put(858,756){\makebox(0,0)[r]{$f_c$ (exp.)}}
\put(902,756){\circle*{12}}
\put(373,637){\circle*{12}}
\put(437,544){\circle*{12}}
\put(475,556){\circle*{12}}
\put(297,739){\circle*{12}}
\put(373,648){\circle*{12}}
\put(475,533){\circle*{12}}
\put(526,442){\circle*{12}}
\put(603,426){\circle*{12}}
\put(794,380){\circle*{12}}
\put(858,711){\makebox(0,0)[r]{simulation}}
\put(902,711){\circle{18}}
\put(284,648){\circle{18}}
\put(316,556){\circle{18}}
\put(348,518){\circle{18}}
\put(379,464){\circle{18}}
\put(411,419){\circle{18}}
\put(475,373){\circle{18}}
\put(603,327){\circle{18}}
\put(730,293){\circle{18}}
\put(794,283){\circle{18}}
\put(858,278){\circle{18}}
\put(985,266){\circle{18}}
\thinlines \path(880,711)(946,711)
\thinlines \path(880,721)(880,701)
\thinlines \path(946,721)(946,701)
\thinlines \path(284,571)(284,724)
\thinlines \path(274,571)(294,571)
\thinlines \path(274,724)(294,724)
\thinlines \path(316,510)(316,602)
\thinlines \path(306,510)(326,510)
\thinlines \path(306,602)(326,602)
\thinlines \path(348,487)(348,548)
\thinlines \path(338,487)(358,487)
\thinlines \path(338,548)(358,548)
\thinlines \path(379,449)(379,480)
\thinlines \path(369,449)(389,449)
\thinlines \path(369,480)(389,480)
\thinlines \path(411,411)(411,426)
\thinlines \path(401,411)(421,411)
\thinlines \path(401,426)(421,426)
\thinlines \path(475,357)(475,388)
\thinlines \path(465,357)(485,357)
\thinlines \path(465,388)(485,388)
\thinlines \path(603,319)(603,335)
\thinlines \path(593,319)(613,319)
\thinlines \path(593,335)(613,335)
\thinlines \path(730,285)(730,300)
\thinlines \path(720,285)(740,285)
\thinlines \path(720,300)(740,300)
\thinlines \path(794,278)(794,287)
\thinlines \path(784,278)(804,278)
\thinlines \path(784,287)(804,287)
\thinlines \path(858,274)(858,282)
\thinlines \path(848,274)(868,274)
\thinlines \path(848,282)(868,282)
\thinlines \path(985,264)(985,267)
\thinlines \path(975,264)(995,264)
\thinlines \path(975,267)(995,267)
\thinlines \path(284,642)(284,642)(291,615)(298,592)(305,571)(312,554)(319,537)(326,523)(333,510)(340,498)(348,487)(355,477)(362,468)(369,460)(376,452)(383,444)(390,437)(397,431)(404,424)(411,419)(418,413)(425,408)(433,403)(440,398)(447,394)(454,389)(461,385)(468,381)(475,378)(482,374)(489,371)(496,367)(503,364)(510,361)(518,358)(525,355)(532,352)(539,350)(546,347)(553,345)(560,342)(567,340)(574,338)(581,335)(588,333)(595,331)(603,329)(610,327)(617,325)(624,323)(631,322)
\thinlines \path(631,322)(638,320)(645,318)(652,316)(659,315)(666,313)(673,312)(680,310)(688,309)(695,307)(702,306)(709,304)(716,303)(723,301)(730,300)(737,299)(744,298)(751,296)(758,295)(765,294)(773,293)(780,292)(787,291)(794,289)(801,288)(808,287)(815,286)(822,285)(829,284)(836,283)(843,282)(850,281)(858,280)(865,279)(872,279)(879,278)(886,277)(893,276)(900,275)(907,274)(914,273)(921,273)(928,272)(935,271)(942,270)(950,269)(957,269)(964,268)(971,267)(978,267)(985,266)
\thinlines \drawline[-95](293,877)(298,853)(305,822)(312,794)(319,769)(326,747)(333,727)(340,709)(348,692)(355,676)(362,662)(369,649)(376,637)(383,625)(390,614)(397,604)(404,595)(411,586)(418,577)(425,569)(433,561)(440,554)(447,547)(454,540)(461,534)(468,528)(475,522)(482,517)(489,511)(496,506)(503,501)(510,496)(518,492)(525,487)(532,483)(539,479)(546,475)(553,471)(560,467)(567,464)(574,460)(581,457)(588,454)(595,450)(603,447)(610,444)(617,441)(624,438)(631,435)(638,433)(645,430)
\thinlines \drawline[-95](645,430)(652,427)(659,425)(666,422)(673,420)(680,418)(688,415)(695,413)(702,411)(709,409)(716,406)(723,404)(730,402)(737,400)(744,398)(751,397)(758,395)(765,393)(773,391)(780,389)(787,388)(794,386)(801,384)(808,383)(815,381)(822,379)(829,378)(836,376)(843,375)(850,373)(858,372)(865,370)(872,369)(879,368)(886,366)(893,365)(900,364)(907,362)(914,361)(921,360)(928,359)(935,357)(942,356)(950,355)(957,354)(964,353)(971,351)(978,350)(985,349)
\end{picture} }
  \hbox{Fig. 1 \hspace{6.5cm} Fig. 2}
  \caption{Sketch of the physical system undergoing horizontal vibrations for 
           (a) low and (b) higher excitations.}
  \label{fig: 1}
\end{figure}
\begin{figure}[ht]
  \vspace*{-9ex}
  \caption{Phase diagram for the solid--fluid transition given in 
    ref.~\protect{\cite{ristow}}: experimental points are $\star\ (f_0)$ and 
    $\bullet\ (f_c)$; data from numerical simulations are $\circ$ 
    with solid line as best fit. Dashed line shows theoretical curve for
    equal size particles.}
  \label{fig: 2}
\end{figure}
\fi

We will study the above mentioned system numerically using spherical particles
which are bound to move in two dimensions and only interact via contact forces.
Periodic boundary conditions are used in the direction of the shaking
(horizontal direction) and 342 particles are placed to form two layers initially. A
predictor-corrector time integration scheme commonly used in molecular dynamics
type simulations is used. The forces acting on particle $i$, having a radius
of $r_i=d_i/2$, during a collision with particle $j$ are in the normal direction 
($\hat{n}$)
\begin{equation}
  F_{ij}^n = - k_n (r_i + r_j - \vec{r}_{ij}\hat{n}) -
               \gamma_n \vec{v}_{ij} \hat{n}
  \label{eq: fn}
\end{equation}
and in the shear direction ($\hat{s}$)
\begin{equation}
  F_{ij}^s = \mbox{sign}(\vec{v}_{ij}\hat{s}) \cdot
             \min(\gamma_s \vec{v}_{ij}\hat{s}, \mu | F_{ij}^n |) \ .
  \label{eq: fs}
\end{equation}
Here $\vec{r}_{ij}$ stands for the vector joining both centers of mass and
$\vec{v}_{ij}$ denotes the relative motion of the two particles. The particles
are allowed to rotate since this is well observed experimentally. The parameter
$k_n$ is related to the Young modulus and is chosen as $10^6$~N/m which is 
high enough to avoid unphysical results due to the {\em detachment 
effect}~\cite{luding1}. The parameter $\gamma_n$ is related to the
experimentally more accessible restitution coefficient $e_n$ and a value of
$e_n=0.75$ was used in all simulations in accordance with the
experiment~\cite{ristow}. The relevant parameters entering the shear force are
given by the detailed collision experiments conducted by Foerster et
al.~\cite{foerster}. Our parameter $\gamma_s$ was set to such a high value that
the shear force was mostly dominated by the Coulomb threshold condition $\mu |
F_{ij}^n |$ where a value of $\mu=0.1$ was used for particle-particle and
$\mu_w=0.13$ for particle-wall contacts. Our force law in the shear direction,
eq.~(\ref{eq: fs}), is a slight oversimplification with respect to the
experimental results in the low shear velocity regime~\cite{foerster} since it
does not contain the reversion of rotation. We checked the validity of this
approximation by using a shear force that exactly matches the data points given
in ref.~\cite{foerster} for glass spheres and even though the transition point
shifted slightly, neither the scaling of the transition points with $\Gamma$
nor the critical point exponents changed their values within the numerical
accuracy. A more detailed discussion of the different force laws was given by
Sch\"afer et al.~\cite{schaefer}. A system size of 10.08 cm was used and the
average particle diameter was $\langle d\rangle = 0.56$ mm with a uniform size
variation of $\pm$ 0.04 mm. The container bottom consists of smaller particles 
with an average size of $\langle d\rangle/4$ denoted by black particles in
fig.~\ref{fig: 1}.

In order to detect the relative motion between particles, we look at the
average kinetic energy of all particles, defined as $e_{\mbox{kin}}(\tau) :=
\frac{1}{2N}\sum_{i=1}^{N} m_i v_i^2(\tau)$, as function of time measured in
the moving reference frame of the channel. This signal is averaged over an
integer number of external vibration cycles, typically 5-10, to give
$E_{\mbox{kin}}$ and we propose that it can be used as an order parameter for
the solid--fluid transition in granular assemblies. In fig.~\ref{fig: 4}a, we
show how  $E_{\mbox{kin}}$ depends on the shaking frequency $f$ measured in Hz,
for two different shaking amplitudes $A_0$ ($\bullet$ -- 7 cm, $\circ$ -- 3
cm).The  errorbars are less than the symbol size. Below a well defined
threshold $f_c$,  $E_{\mbox{kin}}$ is approximately zero which indicates that
the particles are  all at rest and can be viewed as a solid. Above the
threshold, a monotonic,  drastic increase is visible. Particles are moving
around relative to each  other and the average density decreases. The state of
the system in this  regime can be best described by a fluid. 
\begin{figure}[ht]
  { \hspace{-1.5cm}
\ifx\epl\undefined
  \setlength{\unitlength}{0.240900pt}
  \begin{picture}(963,825)(0,0)
\else
  \setlength{\unitlength}{0.225pt}
  \begin{picture}(963,825)(-120,0)
\fi
\thicklines \path(220,113)(240,113)
\thicklines \path(899,113)(879,113)
\put(198,113){\makebox(0,0)[r]{0}}
\thicklines \path(220,251)(240,251)
\thicklines \path(899,251)(879,251)
\put(198,251){\makebox(0,0)[r]{0.1}}
\thicklines \path(220,389)(240,389)
\thicklines \path(899,389)(879,389)
\put(198,389){\makebox(0,0)[r]{0.2}}
\thicklines \path(220,526)(240,526)
\thicklines \path(899,526)(879,526)
\put(198,526){\makebox(0,0)[r]{0.3}}
\thicklines \path(220,664)(240,664)
\thicklines \path(899,664)(879,664)
\put(198,664){\makebox(0,0)[r]{0.4}}
\thicklines \path(220,802)(240,802)
\thicklines \path(899,802)(879,802)
\put(198,802){\makebox(0,0)[r]{0.5}}
\thicklines \path(220,113)(220,133)
\thicklines \path(220,802)(220,782)
\put(220,68){\makebox(0,0){0}}
\thicklines \path(356,113)(356,133)
\thicklines \path(356,802)(356,782)
\put(356,68){\makebox(0,0){1}}
\thicklines \path(492,113)(492,133)
\thicklines \path(492,802)(492,782)
\put(492,68){\makebox(0,0){2}}
\thicklines \path(627,113)(627,133)
\thicklines \path(627,802)(627,782)
\put(627,68){\makebox(0,0){3}}
\thicklines \path(763,113)(763,133)
\thicklines \path(763,802)(763,782)
\put(763,68){\makebox(0,0){4}}
\thicklines \path(899,113)(899,133)
\thicklines \path(899,802)(899,782)
\put(899,68){\makebox(0,0){5}}
\thicklines \path(220,113)(899,113)(899,802)(220,802)(220,113)
\put(45,457){\makebox(0,0)[l]{\shortstack{$E_{\mbox{kin}}$}}}
\put(559,0){\makebox(0,0){$f$ [Hz]}}
\put(763,251){\makebox(0,0){\bf (a)}}
\put(383,733){\makebox(0,0){$A_0\! =\! 7$ cm}}
\put(709,664){\makebox(0,0){$A_0\! =\! 3$ cm}}
\thinlines \path(315,113)(315,113)(329,114)(342,117)(356,134)(369,169)(383,212)(397,258)(410,308)(424,358)(437,413)(451,467)(464,526)(478,587)(492,651)(505,713)(519,788)
\put(315,113){\circle*{12}}
\put(329,114){\circle*{12}}
\put(342,117){\circle*{12}}
\put(356,134){\circle*{12}}
\put(369,169){\circle*{12}}
\put(383,212){\circle*{12}}
\put(397,258){\circle*{12}}
\put(410,308){\circle*{12}}
\put(424,358){\circle*{12}}
\put(437,413){\circle*{12}}
\put(451,467){\circle*{12}}
\put(464,526){\circle*{12}}
\put(478,587){\circle*{12}}
\put(492,651){\circle*{12}}
\put(505,713){\circle*{12}}
\put(519,788){\circle*{12}}
\thinlines \path(356,113)(356,113)(390,114)(407,116)(424,123)(441,132)(458,148)(475,161)(492,178)(509,196)(526,212)(543,231)(560,247)(593,288)(627,330)(661,373)(695,424)(729,478)(763,534)(797,595)(831,657)(865,726)
\put(356,113){\circle{18}}
\put(390,114){\circle{18}}
\put(407,116){\circle{18}}
\put(424,123){\circle{18}}
\put(441,132){\circle{18}}
\put(458,148){\circle{18}}
\put(475,161){\circle{18}}
\put(492,178){\circle{18}}
\put(509,196){\circle{18}}
\put(526,212){\circle{18}}
\put(543,231){\circle{18}}
\put(560,247){\circle{18}}
\put(593,288){\circle{18}}
\put(627,330){\circle{18}}
\put(661,373){\circle{18}}
\put(695,424){\circle{18}}
\put(729,478){\circle{18}}
\put(763,534){\circle{18}}
\put(797,595){\circle{18}}
\put(831,657){\circle{18}}
\put(865,726){\circle{18}}
\end{picture} \hfill \hspace{-0.5cm}
\ifx\epl\undefined
  \setlength{\unitlength}{0.240900pt}
  \begin{picture}(963,825)(0,0)
\else
  \setlength{\unitlength}{0.225pt}
  \begin{picture}(963,825)(-60,0)
\fi
\thicklines \path(220,113)(240,113)
\thicklines \path(899,113)(879,113)
\put(198,113){\makebox(0,0)[r]{0}}
\thicklines \path(220,228)(240,228)
\thicklines \path(899,228)(879,228)
\put(198,228){\makebox(0,0)[r]{}}
\thicklines \path(220,343)(240,343)
\thicklines \path(899,343)(879,343)
\put(198,343){\makebox(0,0)[r]{0.005}}
\thicklines \path(220,458)(240,458)
\thicklines \path(899,458)(879,458)
\put(198,458){\makebox(0,0)[r]{}}
\thicklines \path(220,572)(240,572)
\thicklines \path(899,572)(879,572)
\put(198,572){\makebox(0,0)[r]{0.01}}
\thicklines \path(220,687)(240,687)
\thicklines \path(899,687)(879,687)
\put(198,687){\makebox(0,0)[r]{}}
\thicklines \path(220,802)(240,802)
\thicklines \path(899,802)(879,802)
\put(198,802){\makebox(0,0)[r]{0.015}}
\thicklines \path(220,113)(220,133)
\thicklines \path(220,802)(220,782)
\put(220,68){\makebox(0,0){0}}
\thicklines \path(356,113)(356,133)
\thicklines \path(356,802)(356,782)
\put(356,68){\makebox(0,0){1}}
\thicklines \path(492,113)(492,133)
\thicklines \path(492,802)(492,782)
\put(492,68){\makebox(0,0){2}}
\thicklines \path(627,113)(627,133)
\thicklines \path(627,802)(627,782)
\put(627,68){\makebox(0,0){3}}
\thicklines \path(763,113)(763,133)
\thicklines \path(763,802)(763,782)
\put(763,68){\makebox(0,0){4}}
\thicklines \path(899,113)(899,133)
\thicklines \path(899,802)(899,782)
\put(899,68){\makebox(0,0){5}}
\thicklines \path(220,113)(899,113)(899,802)(220,802)(220,113)
\put(45,457){\makebox(0,0)[l]{\shortstack{$T_{\mbox{gran}}$}}}
\put(559,0){\makebox(0,0){$f$ [Hz]}}
\put(763,251){\makebox(0,0){\bf (b)}}
\put(383,710){\makebox(0,0){$A_0\! =\! 7$ cm}}
\put(722,664){\makebox(0,0){$A_0\! =\! 3$ cm}}
\put(315,116){\circle*{12}}
\put(329,118){\circle*{12}}
\put(342,123){\circle*{12}}
\put(356,149){\circle*{12}}
\put(369,209){\circle*{12}}
\put(383,256){\circle*{12}}
\put(397,301){\circle*{12}}
\put(410,356){\circle*{12}}
\put(424,403){\circle*{12}}
\put(437,449){\circle*{12}}
\put(451,500){\circle*{12}}
\put(464,555){\circle*{12}}
\put(478,600){\circle*{12}}
\put(492,635){\circle*{12}}
\put(505,684){\circle*{12}}
\put(519,742){\circle*{12}}
\thinlines \path(344,113)(350,137)(357,162)(364,187)(371,212)(378,237)(385,261)(391,286)(398,311)(405,336)(412,361)(419,386)(426,410)(433,435)(439,460)(446,485)(453,510)(460,535)(467,559)(474,584)(481,609)(487,634)(494,659)(501,683)(508,708)(515,733)(522,758)(529,783)(534,802)
\put(356,116){\circle{18}}
\put(390,119){\circle{18}}
\put(407,121){\circle{18}}
\put(424,133){\circle{18}}
\put(441,151){\circle{18}}
\put(458,169){\circle{18}}
\put(475,187){\circle{18}}
\put(492,209){\circle{18}}
\put(509,230){\circle{18}}
\put(526,250){\circle{18}}
\put(543,267){\circle{18}}
\put(560,288){\circle{18}}
\put(593,331){\circle{18}}
\put(627,377){\circle{18}}
\put(661,408){\circle{18}}
\put(695,450){\circle{18}}
\put(729,498){\circle{18}}
\put(763,530){\circle{18}}
\put(797,573){\circle{18}}
\put(831,623){\circle{18}}
\put(865,669){\circle{18}}
\thinlines \path(410,113)(412,115)(419,123)(426,132)(433,140)(439,148)(446,156)(453,164)(460,172)(467,180)(474,188)(481,196)(487,205)(494,213)(501,221)(508,229)(515,237)(522,245)(529,253)(535,261)(542,270)(549,278)(556,286)(563,294)(570,302)(577,310)(584,318)(590,326)(597,335)(604,343)(611,351)(618,359)(625,367)(632,375)(638,383)(645,391)(652,399)(659,408)(666,416)(673,424)(680,432)(686,440)(693,448)(700,456)(707,464)(714,473)(721,481)(728,489)(734,497)(741,505)(748,513)
\thinlines \path(748,513)(755,521)(762,529)(769,537)(776,546)(782,554)(789,562)(796,570)(803,578)(810,586)(817,594)(824,602)(830,611)(837,619)(844,627)(851,635)(858,643)(865,651)(872,659)(878,667)(885,676)(892,684)(899,692)
\end{picture} }
  \caption{Averaged kinetic energy (a) and granular temperature (b) as 
           function of frequency $f$ for two different shaking amplitudes 
           $A_0 = 7$ cm and $A_0 = 3$ cm.}
  \label{fig: 4}
\end{figure}

A more accurate method to determine $f_c$ is given by subtracting the bulk
motion from the kinetic energy in the following fashion $t_{\mbox{gran}}(\tau)
:= \frac{1}{2N}\sum_{i=1}^N m_i (v_i(\tau) - \langle v\rangle)^2$ which is
commonly referred to as the granular temperature of the system. Here $\langle
v\rangle$ stands for the average particle velocity at time $\tau$. This
quantity is also averaged over an integer number of full cycles of the external
vibration to give $T_{\mbox{gran}}$. Below the transition point, this quantity
is zero and it is non-zero above it. Since $T_{\mbox{gran}}$ scales linearly
with $f$ for a wide frequency range it can easily be extrapolated to the limit
$T_{\mbox{gran}} \rightarrow 0$ to give a very accurate value of $f_c$. This
procedure is demonstrated in fig.~\ref{fig: 4}b for the two shaking
amplitudes $A_0=7$ cm and $A_0=3$ cm and gives values of $f_c=0.91$ and 1.4 Hz,
respectively. When kinetic theories were applied to vibrated beds the granular
temperature was found to scale as $T \sim (A_0\,\omega)^2$~\cite{goldshtein,lee}
which is in contrast to our finding and indicates that the current theories do
not adequately describe the solid-fluid transition in granular materials.

When the critical frequencies for an amplitude range of 0.5 to 9 cm are plotted 
on a double logarithmic scale, one finds a scaling of $A_0 f_c^2 =
\mbox{const.}$ which is shown in fig.~\ref{fig: 5}. Such a scaling relation was 
also found for the transition point for surface instabilities and the onset of 
convection rolls in a sand pile under {\em vertical} vibrations~\cite{evesque}. 
It can be understood by looking at the force balance along the shear direction 
of one particle on a row of particles. This gives as necessary condition for
particle motion $A_0\,\omega^2/g > \tan\Theta$ where $\Theta$ measures the angle
from the vertical of the center of mass of the top particle to the next
particle center below it. For equal size spheres this gives $\Theta_e=30^\circ$
whereas in our case $\Theta=11.5^\circ$.
\ifx\epl\undefined
\begin{figure}[ht]
  \centerline{ 
\ifx\epl\undefined
  \setlength{\unitlength}{0.240900pt}
  \begin{picture}(963,825)(0,0)
\else
  \setlength{\unitlength}{0.225pt}
  \begin{picture}(963,825)(-120,0)
\fi
\thicklines \path(220,113)(240,113)
\thicklines \path(899,113)(879,113)
\put(198,113){\makebox(0,0)[r]{0.1}}
\thicklines \path(220,217)(230,217)
\thicklines \path(899,217)(889,217)
\put(198,217){\makebox(0,0)[r]{}}
\thicklines \path(220,277)(230,277)
\thicklines \path(899,277)(889,277)
\put(198,277){\makebox(0,0)[r]{}}
\thicklines \path(220,320)(230,320)
\thicklines \path(899,320)(889,320)
\put(198,320){\makebox(0,0)[r]{}}
\thicklines \path(220,354)(230,354)
\thicklines \path(899,354)(889,354)
\put(198,354){\makebox(0,0)[r]{}}
\thicklines \path(220,381)(230,381)
\thicklines \path(899,381)(889,381)
\put(198,381){\makebox(0,0)[r]{}}
\thicklines \path(220,404)(230,404)
\thicklines \path(899,404)(889,404)
\put(198,404){\makebox(0,0)[r]{}}
\thicklines \path(220,424)(230,424)
\thicklines \path(899,424)(889,424)
\put(198,424){\makebox(0,0)[r]{}}
\thicklines \path(220,442)(230,442)
\thicklines \path(899,442)(889,442)
\put(198,442){\makebox(0,0)[r]{}}
\thicklines \path(220,458)(240,458)
\thicklines \path(899,458)(879,458)
\put(198,458){\makebox(0,0)[r]{1}}
\thicklines \path(220,561)(230,561)
\thicklines \path(899,561)(889,561)
\put(198,561){\makebox(0,0)[r]{}}
\thicklines \path(220,622)(230,622)
\thicklines \path(899,622)(889,622)
\put(198,622){\makebox(0,0)[r]{}}
\thicklines \path(220,665)(230,665)
\thicklines \path(899,665)(889,665)
\put(198,665){\makebox(0,0)[r]{}}
\thicklines \path(220,698)(230,698)
\thicklines \path(899,698)(889,698)
\put(198,698){\makebox(0,0)[r]{}}
\thicklines \path(220,726)(230,726)
\thicklines \path(899,726)(889,726)
\put(198,726){\makebox(0,0)[r]{}}
\thicklines \path(220,749)(230,749)
\thicklines \path(899,749)(889,749)
\put(198,749){\makebox(0,0)[r]{}}
\thicklines \path(220,769)(230,769)
\thicklines \path(899,769)(889,769)
\put(198,769){\makebox(0,0)[r]{}}
\thicklines \path(220,786)(230,786)
\thicklines \path(899,786)(889,786)
\put(198,786){\makebox(0,0)[r]{}}
\thicklines \path(220,802)(240,802)
\thicklines \path(899,802)(879,802)
\put(198,802){\makebox(0,0)[r]{10}}
\thicklines \path(220,113)(220,133)
\thicklines \path(220,802)(220,782)
\put(220,68){\makebox(0,0){0.1}}
\thicklines \path(322,113)(322,123)
\thicklines \path(322,802)(322,792)
\put(322,68){\makebox(0,0){}}
\thicklines \path(382,113)(382,123)
\thicklines \path(382,802)(382,792)
\put(382,68){\makebox(0,0){}}
\thicklines \path(424,113)(424,123)
\thicklines \path(424,802)(424,792)
\put(424,68){\makebox(0,0){}}
\thicklines \path(457,113)(457,123)
\thicklines \path(457,802)(457,792)
\put(457,68){\makebox(0,0){}}
\thicklines \path(484,113)(484,123)
\thicklines \path(484,802)(484,792)
\put(484,68){\makebox(0,0){}}
\thicklines \path(507,113)(507,123)
\thicklines \path(507,802)(507,792)
\put(507,68){\makebox(0,0){}}
\thicklines \path(527,113)(527,123)
\thicklines \path(527,802)(527,792)
\put(527,68){\makebox(0,0){}}
\thicklines \path(544,113)(544,123)
\thicklines \path(544,802)(544,792)
\put(544,68){\makebox(0,0){}}
\thicklines \path(560,113)(560,133)
\thicklines \path(560,802)(560,782)
\put(560,68){\makebox(0,0){1}}
\thicklines \path(662,113)(662,123)
\thicklines \path(662,802)(662,792)
\put(662,68){\makebox(0,0){}}
\thicklines \path(721,113)(721,123)
\thicklines \path(721,802)(721,792)
\put(721,68){\makebox(0,0){}}
\thicklines \path(764,113)(764,123)
\thicklines \path(764,802)(764,792)
\put(764,68){\makebox(0,0){}}
\thicklines \path(797,113)(797,123)
\thicklines \path(797,802)(797,792)
\put(797,68){\makebox(0,0){}}
\thicklines \path(824,113)(824,123)
\thicklines \path(824,802)(824,792)
\put(824,68){\makebox(0,0){}}
\thicklines \path(846,113)(846,123)
\thicklines \path(846,802)(846,792)
\put(846,68){\makebox(0,0){}}
\thicklines \path(866,113)(866,123)
\thicklines \path(866,802)(866,792)
\put(866,68){\makebox(0,0){}}
\thicklines \path(883,113)(883,123)
\thicklines \path(883,802)(883,792)
\put(883,68){\makebox(0,0){}}
\thicklines \path(899,113)(899,133)
\thicklines \path(899,802)(899,782)
\put(899,68){\makebox(0,0){10}}
\thicklines \path(220,113)(899,113)(899,802)(220,802)(220,113)
\put(45,457){\makebox(0,0)[l]{\shortstack{$f_c$ \\ \protect{[Hz]}}}}
\put(559,-22){\makebox(0,0){$A_0$ [cm]}}
\put(457,645){\circle{18}}
\put(517,617){\circle{18}}
\put(560,603){\circle{18}}
\put(592,582){\circle{18}}
\put(619,561){\circle{18}}
\put(662,537){\circle{18}}
\put(721,508){\circle{18}}
\put(764,482){\circle{18}}
\put(781,473){\circle{18}}
\put(797,469){\circle{18}}
\put(824,458){\circle{18}}
\put(846,443){\circle{18}}
\put(883,420){\circle{18}}
\thinlines \path(457,622)(457,665)
\thinlines \path(447,622)(467,622)
\thinlines \path(447,665)(467,665)
\thinlines \path(517,600)(517,632)
\thinlines \path(507,600)(527,600)
\thinlines \path(507,632)(527,632)
\thinlines \path(560,592)(560,614)
\thinlines \path(550,592)(570,592)
\thinlines \path(550,614)(570,614)
\thinlines \path(592,575)(592,588)
\thinlines \path(582,575)(602,575)
\thinlines \path(582,588)(602,588)
\thinlines \path(619,557)(619,565)
\thinlines \path(609,557)(629,557)
\thinlines \path(609,565)(629,565)
\thinlines \path(662,528)(662,545)
\thinlines \path(652,528)(672,528)
\thinlines \path(652,545)(672,545)
\thinlines \path(721,502)(721,513)
\thinlines \path(711,502)(731,502)
\thinlines \path(711,513)(731,513)
\thinlines \path(764,475)(764,488)
\thinlines \path(754,475)(774,475)
\thinlines \path(754,488)(774,488)
\thinlines \path(781,469)(781,477)
\thinlines \path(771,469)(791,469)
\thinlines \path(771,477)(791,477)
\thinlines \path(797,466)(797,472)
\thinlines \path(787,466)(807,466)
\thinlines \path(787,472)(807,472)
\thinlines \path(824,456)(824,459)
\thinlines \path(814,456)(834,456)
\thinlines \path(814,459)(834,459)
\thinlines \path(846,442)(846,445)
\thinlines \path(836,442)(856,442)
\thinlines \path(836,445)(856,445)
\thinlines \path(883,418)(883,422)
\thinlines \path(873,418)(893,418)
\thinlines \path(873,422)(893,422)
\thinlines \path(457,644)(457,644)(462,642)(466,640)(470,637)(475,635)(479,633)(483,631)(487,629)(492,627)(496,624)(500,622)(505,620)(509,618)(513,616)(518,613)(522,611)(526,609)(530,607)(535,605)(539,603)(543,600)(548,598)(552,596)(556,594)(561,592)(565,589)(569,587)(574,585)(578,583)(582,581)(586,579)(591,576)(595,574)(599,572)(604,570)(608,568)(612,565)(617,563)(621,561)(625,559)(629,557)(634,554)(638,552)(642,550)(647,548)(651,546)(655,544)(660,541)(664,539)(668,537)
\thinlines \path(668,537)(673,535)(677,533)(681,530)(685,528)(690,526)(694,524)(698,522)(703,520)(707,517)(711,515)(716,513)(720,511)(724,509)(728,506)(733,504)(737,502)(741,500)(746,498)(750,496)(754,493)(759,491)(763,489)(767,487)(772,485)(776,482)(780,480)(784,478)(789,476)(793,474)(797,471)(802,469)(806,467)(810,465)(815,463)(819,461)(823,458)(828,456)(832,454)(836,452)(840,450)(845,447)(849,445)(853,443)(858,441)(862,439)(866,437)(871,434)(875,432)(879,430)(883,428)
\end{picture} }
  \caption{Critical frequency as function of the external shaking amplitude
           $A_0$}
  \label{fig: 5}
\end{figure}
\fi

Knowing that the transition point scales as $A_0 f_c^2 = \mbox{const.}$, we
replot in fig.~\ref{fig: 6} the averaged kinetic energy $E_{\mbox{kin}}$
divided by $A_0$ as function of the dimensionless quantity $\Gamma' := A_0
f^2/(g \tan\Theta)$. All data points collapse onto one straight line, giving a 
transition point of $\Gamma_c' \approx 1.07$. For the vertical shaking
experiments, a transition point of  $\Gamma_c \approx 1.2$ was
found~\cite{evesque}.
\ifx\epl\undefined
\begin{figure}[ht]
  \centerline{ 
\ifx\epl\undefined
  \setlength{\unitlength}{0.240900pt}
  \begin{picture}(963,825)(0,0)
\else
  \setlength{\unitlength}{0.225pt}
  \begin{picture}(963,825)(-60,55)
\fi
\thicklines \path(221,179)(241,179)
\thicklines \path(900,179)(880,179)
\put(199,179){\makebox(0,0)[r]{0}}
\thicklines \path(221,348)(241,348)
\thicklines \path(900,348)(880,348)
\put(199,348){\makebox(0,0)[r]{0.05}}
\thicklines \path(221,517)(241,517)
\thicklines \path(900,517)(880,517)
\put(199,517){\makebox(0,0)[r]{0.1}}
\thicklines \path(221,686)(241,686)
\thicklines \path(900,686)(880,686)
\put(199,686){\makebox(0,0)[r]{0.15}}
\thicklines \path(221,855)(241,855)
\thicklines \path(900,855)(880,855)
\put(199,855){\makebox(0,0)[r]{0.2}}
\thicklines \path(221,179)(221,199)
\thicklines \path(221,855)(221,835)
\put(221,134){\makebox(0,0){0}}
\thicklines \path(266,179)(266,199)
\thicklines \path(266,855)(266,835)
\put(266,134){\makebox(0,0){}}
\thicklines \path(312,179)(312,199)
\thicklines \path(312,855)(312,835)
\put(312,134){\makebox(0,0){2}}
\thicklines \path(357,179)(357,199)
\thicklines \path(357,855)(357,835)
\put(357,134){\makebox(0,0){}}
\thicklines \path(402,179)(402,199)
\thicklines \path(402,855)(402,835)
\put(402,134){\makebox(0,0){4}}
\thicklines \path(447,179)(447,199)
\thicklines \path(447,855)(447,835)
\put(447,134){\makebox(0,0){}}
\thicklines \path(493,179)(493,199)
\thicklines \path(493,855)(493,835)
\put(493,134){\makebox(0,0){6}}
\thicklines \path(538,179)(538,199)
\thicklines \path(538,855)(538,835)
\put(538,134){\makebox(0,0){}}
\thicklines \path(583,179)(583,199)
\thicklines \path(583,855)(583,835)
\put(583,134){\makebox(0,0){8}}
\thicklines \path(628,179)(628,199)
\thicklines \path(628,855)(628,835)
\put(628,134){\makebox(0,0){}}
\thicklines \path(674,179)(674,199)
\thicklines \path(674,855)(674,835)
\put(674,134){\makebox(0,0){10}}
\thicklines \path(719,179)(719,199)
\thicklines \path(719,855)(719,835)
\put(719,134){\makebox(0,0){}}
\thicklines \path(764,179)(764,199)
\thicklines \path(764,855)(764,835)
\put(764,134){\makebox(0,0){12}}
\thicklines \path(809,179)(809,199)
\thicklines \path(809,855)(809,835)
\put(809,134){\makebox(0,0){}}
\thicklines \path(855,179)(855,199)
\thicklines \path(855,855)(855,835)
\put(855,134){\makebox(0,0){14}}
\thicklines \path(900,179)(900,199)
\thicklines \path(900,855)(900,835)
\put(900,134){\makebox(0,0){}}
\thicklines \path(221,179)(900,179)(900,855)(221,855)(221,179)
\put(0,602){\makebox(0,0)[l]{\shortstack{$E_{\mbox{kin}}/A_0$}}}
\put(560,40){\makebox(0,0){$\Gamma'$}}
\put(742,365){\makebox(0,0)[r]{$A_0\! =\! 7$ cm}}
\put(252,179){\circle*{12}}
\put(261,179){\circle*{12}}
\put(272,180){\circle*{12}}
\put(284,186){\circle*{12}}
\put(297,199){\circle*{12}}
\put(311,214){\circle*{12}}
\put(327,230){\circle*{12}}
\put(344,247){\circle*{12}}
\put(362,265){\circle*{12}}
\put(381,284){\circle*{12}}
\put(402,303){\circle*{12}}
\put(424,324){\circle*{12}}
\put(447,345){\circle*{12}}
\put(471,368){\circle*{12}}
\put(497,389){\circle*{12}}
\put(524,416){\circle*{12}}
\put(818,365){\circle*{12}}
\put(742,320){\makebox(0,0)[r]{$3$ cm}}
\put(248,179){\circle{18}}
\put(263,180){\circle{18}}
\put(272,181){\circle{18}}
\put(281,187){\circle{18}}
\put(292,195){\circle{18}}
\put(303,207){\circle{18}}
\put(315,218){\circle{18}}
\put(328,232){\circle{18}}
\put(342,247){\circle{18}}
\put(357,260){\circle{18}}
\put(372,275){\circle{18}}
\put(388,288){\circle{18}}
\put(424,322){\circle{18}}
\put(462,356){\circle{18}}
\put(504,392){\circle{18}}
\put(549,434){\circle{18}}
\put(598,478){\circle{18}}
\put(650,523){\circle{18}}
\put(705,573){\circle{18}}
\put(763,624){\circle{18}}
\put(825,680){\circle{18}}
\put(891,731){\circle{18}}
\put(818,320){\circle{18}}
\put(742,275){\makebox(0,0)[r]{$1$ cm}}
\put(230,179){\makebox(0,0){$\star$}}
\put(241,179){\makebox(0,0){$\star$}}
\put(257,180){\makebox(0,0){$\star$}}
\put(277,189){\makebox(0,0){$\star$}}
\put(301,208){\makebox(0,0){$\star$}}
\put(330,236){\makebox(0,0){$\star$}}
\put(364,268){\makebox(0,0){$\star$}}
\put(402,303){\makebox(0,0){$\star$}}
\put(444,343){\makebox(0,0){$\star$}}
\put(491,384){\makebox(0,0){$\star$}}
\put(542,433){\makebox(0,0){$\star$}}
\put(598,480){\makebox(0,0){$\star$}}
\put(659,534){\makebox(0,0){$\star$}}
\put(723,590){\makebox(0,0){$\star$}}
\put(793,652){\makebox(0,0){$\star$}}
\put(866,715){\makebox(0,0){$\star$}}
\put(818,275){\makebox(0,0){$\star$}}
\put(742,230){\makebox(0,0)[r]{fit}}
\thinlines \path(764,230)(872,230)


\path(269,179)(900,758)

\end{picture} }
  \caption{Rescaled averaged kinetic energy as function of dimensionless 
           quantity $\Gamma' := A_0 f^2/(g \tan\Theta)$.}
  \label{fig: 6}
\end{figure}
\else
\fi

The transition point does not depend on the choice of the restitution
coefficient in the normal direction, even though it changes the kinetic energy
and has a dramatic effect on the absolute value of the granular temperature. On
the other hand, changing the Coulomb threshold $\mu$ shifts the critical point
when a high enough  value of $\gamma_s$ is used. For $\mu=0$, the transition
frequency is lowered by roughly 0.5 Hz; whereas increasing it to $\mu=1$
increases $f_c$ by up to 1 Hz as well. Changing the system size to 7.56 or
15.12 cm did neither change the transition point nor the absolute values of the
averaged kinetic energy and granular temperature.
\ifx\epl\undefined
\else
\begin{figure}[ht]
  { \hspace{-1.5cm}
\ifx\epl\undefined
  \setlength{\unitlength}{0.240900pt}
  \begin{picture}(963,825)(0,0)
\else
  \setlength{\unitlength}{0.225pt}
  \begin{picture}(963,825)(-120,0)
\fi
\thicklines \path(220,113)(240,113)
\thicklines \path(899,113)(879,113)
\put(198,113){\makebox(0,0)[r]{0.1}}
\thicklines \path(220,217)(230,217)
\thicklines \path(899,217)(889,217)
\put(198,217){\makebox(0,0)[r]{}}
\thicklines \path(220,277)(230,277)
\thicklines \path(899,277)(889,277)
\put(198,277){\makebox(0,0)[r]{}}
\thicklines \path(220,320)(230,320)
\thicklines \path(899,320)(889,320)
\put(198,320){\makebox(0,0)[r]{}}
\thicklines \path(220,354)(230,354)
\thicklines \path(899,354)(889,354)
\put(198,354){\makebox(0,0)[r]{}}
\thicklines \path(220,381)(230,381)
\thicklines \path(899,381)(889,381)
\put(198,381){\makebox(0,0)[r]{}}
\thicklines \path(220,404)(230,404)
\thicklines \path(899,404)(889,404)
\put(198,404){\makebox(0,0)[r]{}}
\thicklines \path(220,424)(230,424)
\thicklines \path(899,424)(889,424)
\put(198,424){\makebox(0,0)[r]{}}
\thicklines \path(220,442)(230,442)
\thicklines \path(899,442)(889,442)
\put(198,442){\makebox(0,0)[r]{}}
\thicklines \path(220,458)(240,458)
\thicklines \path(899,458)(879,458)
\put(198,458){\makebox(0,0)[r]{1}}
\thicklines \path(220,561)(230,561)
\thicklines \path(899,561)(889,561)
\put(198,561){\makebox(0,0)[r]{}}
\thicklines \path(220,622)(230,622)
\thicklines \path(899,622)(889,622)
\put(198,622){\makebox(0,0)[r]{}}
\thicklines \path(220,665)(230,665)
\thicklines \path(899,665)(889,665)
\put(198,665){\makebox(0,0)[r]{}}
\thicklines \path(220,698)(230,698)
\thicklines \path(899,698)(889,698)
\put(198,698){\makebox(0,0)[r]{}}
\thicklines \path(220,726)(230,726)
\thicklines \path(899,726)(889,726)
\put(198,726){\makebox(0,0)[r]{}}
\thicklines \path(220,749)(230,749)
\thicklines \path(899,749)(889,749)
\put(198,749){\makebox(0,0)[r]{}}
\thicklines \path(220,769)(230,769)
\thicklines \path(899,769)(889,769)
\put(198,769){\makebox(0,0)[r]{}}
\thicklines \path(220,786)(230,786)
\thicklines \path(899,786)(889,786)
\put(198,786){\makebox(0,0)[r]{}}
\thicklines \path(220,802)(240,802)
\thicklines \path(899,802)(879,802)
\put(198,802){\makebox(0,0)[r]{10}}
\thicklines \path(220,113)(220,133)
\thicklines \path(220,802)(220,782)
\put(220,68){\makebox(0,0){0.1}}
\thicklines \path(322,113)(322,123)
\thicklines \path(322,802)(322,792)
\put(322,68){\makebox(0,0){}}
\thicklines \path(382,113)(382,123)
\thicklines \path(382,802)(382,792)
\put(382,68){\makebox(0,0){}}
\thicklines \path(424,113)(424,123)
\thicklines \path(424,802)(424,792)
\put(424,68){\makebox(0,0){}}
\thicklines \path(457,113)(457,123)
\thicklines \path(457,802)(457,792)
\put(457,68){\makebox(0,0){}}
\thicklines \path(484,113)(484,123)
\thicklines \path(484,802)(484,792)
\put(484,68){\makebox(0,0){}}
\thicklines \path(507,113)(507,123)
\thicklines \path(507,802)(507,792)
\put(507,68){\makebox(0,0){}}
\thicklines \path(527,113)(527,123)
\thicklines \path(527,802)(527,792)
\put(527,68){\makebox(0,0){}}
\thicklines \path(544,113)(544,123)
\thicklines \path(544,802)(544,792)
\put(544,68){\makebox(0,0){}}
\thicklines \path(560,113)(560,133)
\thicklines \path(560,802)(560,782)
\put(560,68){\makebox(0,0){1}}
\thicklines \path(662,113)(662,123)
\thicklines \path(662,802)(662,792)
\put(662,68){\makebox(0,0){}}
\thicklines \path(721,113)(721,123)
\thicklines \path(721,802)(721,792)
\put(721,68){\makebox(0,0){}}
\thicklines \path(764,113)(764,123)
\thicklines \path(764,802)(764,792)
\put(764,68){\makebox(0,0){}}
\thicklines \path(797,113)(797,123)
\thicklines \path(797,802)(797,792)
\put(797,68){\makebox(0,0){}}
\thicklines \path(824,113)(824,123)
\thicklines \path(824,802)(824,792)
\put(824,68){\makebox(0,0){}}
\thicklines \path(846,113)(846,123)
\thicklines \path(846,802)(846,792)
\put(846,68){\makebox(0,0){}}
\thicklines \path(866,113)(866,123)
\thicklines \path(866,802)(866,792)
\put(866,68){\makebox(0,0){}}
\thicklines \path(883,113)(883,123)
\thicklines \path(883,802)(883,792)
\put(883,68){\makebox(0,0){}}
\thicklines \path(899,113)(899,133)
\thicklines \path(899,802)(899,782)
\put(899,68){\makebox(0,0){10}}
\thicklines \path(220,113)(899,113)(899,802)(220,802)(220,113)
\put(45,457){\makebox(0,0)[l]{\shortstack{$f_c$ \\ \protect{[Hz]}}}}
\put(559,-22){\makebox(0,0){$A_0$ [cm]}}
\put(457,645){\circle{18}}
\put(517,617){\circle{18}}
\put(560,603){\circle{18}}
\put(592,582){\circle{18}}
\put(619,561){\circle{18}}
\put(662,537){\circle{18}}
\put(721,508){\circle{18}}
\put(764,482){\circle{18}}
\put(781,473){\circle{18}}
\put(797,469){\circle{18}}
\put(824,458){\circle{18}}
\put(846,443){\circle{18}}
\put(883,420){\circle{18}}
\thinlines \path(457,622)(457,665)
\thinlines \path(447,622)(467,622)
\thinlines \path(447,665)(467,665)
\thinlines \path(517,600)(517,632)
\thinlines \path(507,600)(527,600)
\thinlines \path(507,632)(527,632)
\thinlines \path(560,592)(560,614)
\thinlines \path(550,592)(570,592)
\thinlines \path(550,614)(570,614)
\thinlines \path(592,575)(592,588)
\thinlines \path(582,575)(602,575)
\thinlines \path(582,588)(602,588)
\thinlines \path(619,557)(619,565)
\thinlines \path(609,557)(629,557)
\thinlines \path(609,565)(629,565)
\thinlines \path(662,528)(662,545)
\thinlines \path(652,528)(672,528)
\thinlines \path(652,545)(672,545)
\thinlines \path(721,502)(721,513)
\thinlines \path(711,502)(731,502)
\thinlines \path(711,513)(731,513)
\thinlines \path(764,475)(764,488)
\thinlines \path(754,475)(774,475)
\thinlines \path(754,488)(774,488)
\thinlines \path(781,469)(781,477)
\thinlines \path(771,469)(791,469)
\thinlines \path(771,477)(791,477)
\thinlines \path(797,466)(797,472)
\thinlines \path(787,466)(807,466)
\thinlines \path(787,472)(807,472)
\thinlines \path(824,456)(824,459)
\thinlines \path(814,456)(834,456)
\thinlines \path(814,459)(834,459)
\thinlines \path(846,442)(846,445)
\thinlines \path(836,442)(856,442)
\thinlines \path(836,445)(856,445)
\thinlines \path(883,418)(883,422)
\thinlines \path(873,418)(893,418)
\thinlines \path(873,422)(893,422)
\thinlines \path(457,644)(457,644)(462,642)(466,640)(470,637)(475,635)(479,633)(483,631)(487,629)(492,627)(496,624)(500,622)(505,620)(509,618)(513,616)(518,613)(522,611)(526,609)(530,607)(535,605)(539,603)(543,600)(548,598)(552,596)(556,594)(561,592)(565,589)(569,587)(574,585)(578,583)(582,581)(586,579)(591,576)(595,574)(599,572)(604,570)(608,568)(612,565)(617,563)(621,561)(625,559)(629,557)(634,554)(638,552)(642,550)(647,548)(651,546)(655,544)(660,541)(664,539)(668,537)
\thinlines \path(668,537)(673,535)(677,533)(681,530)(685,528)(690,526)(694,524)(698,522)(703,520)(707,517)(711,515)(716,513)(720,511)(724,509)(728,506)(733,504)(737,502)(741,500)(746,498)(750,496)(754,493)(759,491)(763,489)(767,487)(772,485)(776,482)(780,480)(784,478)(789,476)(793,474)(797,471)(802,469)(806,467)(810,465)(815,463)(819,461)(823,458)(828,456)(832,454)(836,452)(840,450)(845,447)(849,445)(853,443)(858,441)(862,439)(866,437)(871,434)(875,432)(879,430)(883,428)
\end{picture} \hfill \hspace{-0.5cm}
\ifx\epl\undefined
  \setlength{\unitlength}{0.240900pt}
  \begin{picture}(963,825)(0,0)
\else
  \setlength{\unitlength}{0.225pt}
  \begin{picture}(963,825)(-60,55)
\fi
\thicklines \path(221,179)(241,179)
\thicklines \path(900,179)(880,179)
\put(199,179){\makebox(0,0)[r]{0}}
\thicklines \path(221,348)(241,348)
\thicklines \path(900,348)(880,348)
\put(199,348){\makebox(0,0)[r]{0.05}}
\thicklines \path(221,517)(241,517)
\thicklines \path(900,517)(880,517)
\put(199,517){\makebox(0,0)[r]{0.1}}
\thicklines \path(221,686)(241,686)
\thicklines \path(900,686)(880,686)
\put(199,686){\makebox(0,0)[r]{0.15}}
\thicklines \path(221,855)(241,855)
\thicklines \path(900,855)(880,855)
\put(199,855){\makebox(0,0)[r]{0.2}}
\thicklines \path(221,179)(221,199)
\thicklines \path(221,855)(221,835)
\put(221,134){\makebox(0,0){0}}
\thicklines \path(266,179)(266,199)
\thicklines \path(266,855)(266,835)
\put(266,134){\makebox(0,0){}}
\thicklines \path(312,179)(312,199)
\thicklines \path(312,855)(312,835)
\put(312,134){\makebox(0,0){2}}
\thicklines \path(357,179)(357,199)
\thicklines \path(357,855)(357,835)
\put(357,134){\makebox(0,0){}}
\thicklines \path(402,179)(402,199)
\thicklines \path(402,855)(402,835)
\put(402,134){\makebox(0,0){4}}
\thicklines \path(447,179)(447,199)
\thicklines \path(447,855)(447,835)
\put(447,134){\makebox(0,0){}}
\thicklines \path(493,179)(493,199)
\thicklines \path(493,855)(493,835)
\put(493,134){\makebox(0,0){6}}
\thicklines \path(538,179)(538,199)
\thicklines \path(538,855)(538,835)
\put(538,134){\makebox(0,0){}}
\thicklines \path(583,179)(583,199)
\thicklines \path(583,855)(583,835)
\put(583,134){\makebox(0,0){8}}
\thicklines \path(628,179)(628,199)
\thicklines \path(628,855)(628,835)
\put(628,134){\makebox(0,0){}}
\thicklines \path(674,179)(674,199)
\thicklines \path(674,855)(674,835)
\put(674,134){\makebox(0,0){10}}
\thicklines \path(719,179)(719,199)
\thicklines \path(719,855)(719,835)
\put(719,134){\makebox(0,0){}}
\thicklines \path(764,179)(764,199)
\thicklines \path(764,855)(764,835)
\put(764,134){\makebox(0,0){12}}
\thicklines \path(809,179)(809,199)
\thicklines \path(809,855)(809,835)
\put(809,134){\makebox(0,0){}}
\thicklines \path(855,179)(855,199)
\thicklines \path(855,855)(855,835)
\put(855,134){\makebox(0,0){14}}
\thicklines \path(900,179)(900,199)
\thicklines \path(900,855)(900,835)
\put(900,134){\makebox(0,0){}}
\thicklines \path(221,179)(900,179)(900,855)(221,855)(221,179)
\put(0,602){\makebox(0,0)[l]{\shortstack{$E_{\mbox{kin}}/A_0$}}}
\put(560,40){\makebox(0,0){$\Gamma'$}}
\put(742,365){\makebox(0,0)[r]{$A_0\! =\! 7$ cm}}
\put(252,179){\circle*{12}}
\put(261,179){\circle*{12}}
\put(272,180){\circle*{12}}
\put(284,186){\circle*{12}}
\put(297,199){\circle*{12}}
\put(311,214){\circle*{12}}
\put(327,230){\circle*{12}}
\put(344,247){\circle*{12}}
\put(362,265){\circle*{12}}
\put(381,284){\circle*{12}}
\put(402,303){\circle*{12}}
\put(424,324){\circle*{12}}
\put(447,345){\circle*{12}}
\put(471,368){\circle*{12}}
\put(497,389){\circle*{12}}
\put(524,416){\circle*{12}}
\put(818,365){\circle*{12}}
\put(742,320){\makebox(0,0)[r]{$3$ cm}}
\put(248,179){\circle{18}}
\put(263,180){\circle{18}}
\put(272,181){\circle{18}}
\put(281,187){\circle{18}}
\put(292,195){\circle{18}}
\put(303,207){\circle{18}}
\put(315,218){\circle{18}}
\put(328,232){\circle{18}}
\put(342,247){\circle{18}}
\put(357,260){\circle{18}}
\put(372,275){\circle{18}}
\put(388,288){\circle{18}}
\put(424,322){\circle{18}}
\put(462,356){\circle{18}}
\put(504,392){\circle{18}}
\put(549,434){\circle{18}}
\put(598,478){\circle{18}}
\put(650,523){\circle{18}}
\put(705,573){\circle{18}}
\put(763,624){\circle{18}}
\put(825,680){\circle{18}}
\put(891,731){\circle{18}}
\put(818,320){\circle{18}}
\put(742,275){\makebox(0,0)[r]{$1$ cm}}
\put(230,179){\makebox(0,0){$\star$}}
\put(241,179){\makebox(0,0){$\star$}}
\put(257,180){\makebox(0,0){$\star$}}
\put(277,189){\makebox(0,0){$\star$}}
\put(301,208){\makebox(0,0){$\star$}}
\put(330,236){\makebox(0,0){$\star$}}
\put(364,268){\makebox(0,0){$\star$}}
\put(402,303){\makebox(0,0){$\star$}}
\put(444,343){\makebox(0,0){$\star$}}
\put(491,384){\makebox(0,0){$\star$}}
\put(542,433){\makebox(0,0){$\star$}}
\put(598,480){\makebox(0,0){$\star$}}
\put(659,534){\makebox(0,0){$\star$}}
\put(723,590){\makebox(0,0){$\star$}}
\put(793,652){\makebox(0,0){$\star$}}
\put(866,715){\makebox(0,0){$\star$}}
\put(818,275){\makebox(0,0){$\star$}}
\put(742,230){\makebox(0,0)[r]{fit}}
\thinlines \path(764,230)(872,230)


\path(269,179)(900,758)

\end{picture} }
  \hbox{Fig. 4 \hspace{6.5cm} Fig. 5}
  \caption{Critical frequency as function of the external shaking amplitude
           $A_0$}
  \label{fig: 5}
\end{figure}
\begin{figure}[ht]
  \vspace*{-3ex}
  \caption{Rescaled averaged kinetic energy as function of dimensionless 
           quantity $\Gamma' := A_0 f^2/(g \tan\Theta)$.}
  \label{fig: 6}
\end{figure}
\fi
We will now turn to the main issue of this letter, namely the calculation of
critical-point exponents for the solid--fluid phase transition in granular
material. The sharp transition at $f_c$ from a zero value kinetic energy (or
granular temperature) to a monotonically increasing value with increasing
frequency suggests that the solid--fluid phase transition in a granular
material undergoing vertical vibrations can be described by a second-order
phase transition. This is stressed by the fact that no hysteresis can be found
in the numerical simulations when increasing or decreasing the driving
frequency, $f$. It was shown above that the kinetic energy or the granular
temperature can both equally well be used as order parameter to characterize
this transition. We use 
\[ 
  \epsilon := \frac{f - f_c}{f_c}
\]
as dimensionless variable to describe the behavior near the critical point
$f_c$~\cite{stanley}. The critical exponent $\lambda$ is then given by
\[ 
  \lambda := \lim_{\epsilon\rightarrow 0} \frac{\ln h(\epsilon)}{\ln \epsilon}
\]
where $h(\epsilon)$ stands for either the kinetic energy $E_{\mbox{kin}}$ or
the granular temperature $T_{\mbox{gran}}$. Assuming the existence of this
limit, we can approximate $h(\epsilon)$ in the vicinity above the critical
point by
\[ 
  h(\epsilon) \sim \epsilon^{\lambda} \ . 
\]
We applied this technique to 13 different shaking amplitudes $A_0$ in the range
0.5 to 9.0 cm and found an universal exponent for the scaling of the kinetic
energy of 1.25 $\pm$ 0.15 and of the granular temperature of 1.0 $\pm$ 0.15.
The error bars stem from the possible error in the critical frequency. But no
systematic trend of the exponent for decreasing or increasing values of $A_0$
can be observed.  In  fig.~\ref{fig: 7}, we show this scaling for shaking
amplitudes of $A_0 = 7, 3$ and 1 cm using either the kinetic energy and an
exponent of 1.25 (a) or the granular temperature and an exponent of 1.0 (b) for
all three curves. For each curve, we used values of $\epsilon$ ranging over at
least one order of magnitude. For $A_0=3$ cm, we extended our data points to
very high excitation amplitudes and find a different exponent for values of
$\epsilon>3$ which might indicate that the transition from the fluid to a more
gas-like state takes place in this regime~\cite{goldshtein}. For the lowest
shaking amplitude, the left most point shows a slight deviation from the fit
which can be attributed to a less accurate value for the transition frequency
for this excitation amplitude and finite size effects.
\begin{figure}[htb]
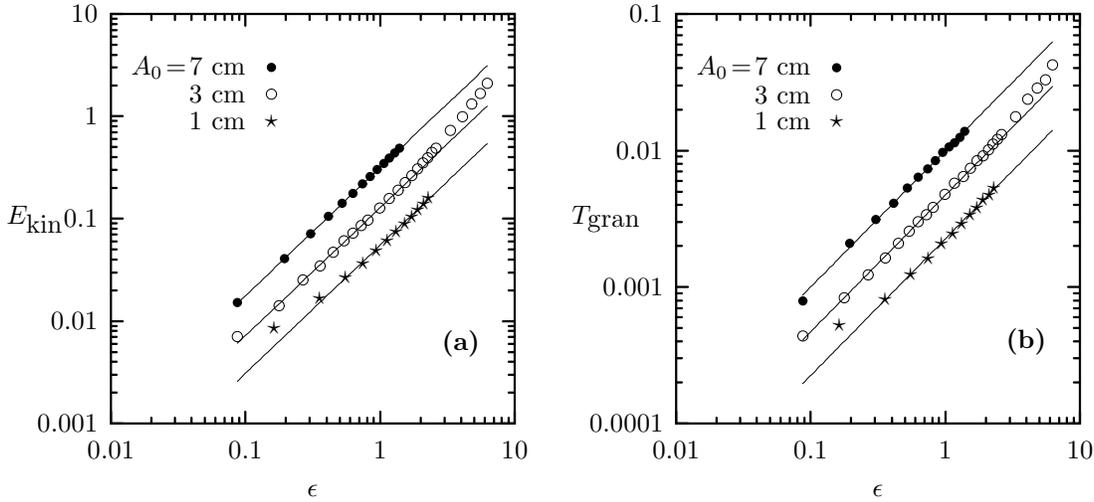

  { \hspace{-1.5cm}\input crit_f7a \hfill \hspace{-0.5cm}\input crit_f7b }
  \caption{Scaling of the kinetic energy (a) and the granular temperature (b)
           as function of dimensionless parameter $\epsilon := (f-f_c)/f_c$. 
           The universal exponent $\lambda$ is 1.25 $\pm$ 0.15 in case (a) 
           and 1.0 $\pm$ 0.15 in case (b).}
  \label{fig: 7}
\end{figure}

In this letter, we investigated numerically the solid--fluid phase transition
of  a granular material exposed to horizontal shaking. By looking at the
averaged kinetic energy or the granular temperature of all particles, we found
that these quantities are only non-zero above a well defined transition
frequency which we called critical frequency and showed no hysteresis. This
suggests that this phase transition is of second-order and we presented the
correct scaling of the transition frequency with respect to the shaking
amplitude. In the vicinity of the critical point, we calculated the critical
exponents associated with the kinetic energy and the granular temperature and
found an universal exponent of 1.25 for the first and of 1.0 for the latter
quantity. The same technique can probably be applied to the solid--fluid phase
transition under {\em vertical} vibration and it would be interesting to
compare the exponents with our results.

\stars

We would like to thank I.~Rehberg, G.~Stra\ss{}burger, S.~Grossmann, A.~Esser 
and K.~Drese for enlightening discussions. This work was supported by the Deutsche 
Forschungsgemeinschaft through Ri 826/1-1.


\end{document}